\documentclass{article}

\usepackage[english]{babel}
\usepackage{amsmath}
\usepackage{amssymb}
\usepackage{graphicx}
\usepackage{authblk}
\usepackage[compress]{cite}
\usepackage{booktabs}

\title{Internal magnetic field distribution in plasmas}
\author{B.\,I.~Lev}
\author{V.\,B.~Tymchyshyn}
\author{A.\,G.~Zagorodny}
\affil{Bogolyubov Institute for Theoretical Physics, National Academy of Sciences, Metrolohichna St. 14-b, Kyiv 03680, Ukraine}

\begin{document}

\maketitle
\begin{abstract}
We calculate stationary probability distribution of magnetic field, generated by moving charges of plasma environment, and stationary probability distribution of force, acting on charged particle in this environment, with magnetic interaction taken into account.
While the former happens to be the Holtsmark distribution, the latter is a modified Holtsmark distribution.
In contrast to prior works, we did no assumptions on the velocity distribution function and thus obtained results should be applicable to wider spectrum of models.
Presented results can be experimentally verified through studies of Zeeman effect or movement of small charged Brownian particles in plasma.
\end{abstract}

\section*{Introduction}

First works on stochastic diffusion in plasma date back to the first half of XX century.
For example, Taylor \cite{bib_taylor} considers diffusion of plasma in presence of external magnetic field with emerging fluctuations of electric field taken into account.
Similar problem of Brownian motion in magnetic field was presented in \cite{bib_kursunoglu} by Kur{\c{s}}uno{\v{g}}lu.
This work was an extension of \cite{bib_chandra} by Chandrasekhar, where probability distribution function of Brownian particle associated with the magnitude of the velocity was calculated.
Recent interest to this problem reemerged with works \cite{bib_czopnik,bib_simoes,bib_hayakawa,bib_ferrari,bib_jimenez,bib_zagorodny,bib_holod,bib_lev_1,bib_lev_2,bib_lev_3,bib_lev_4}.
They take into account electrostatic interaction between particles and interactions between particle and external magnetic field; the latter two \cite{bib_zagorodny,bib_holod} generalize Chandrasekhar approach and calculate correlation function of charge density.

Advances in spectroscopy led to new insights into electric and magnetic fields acting on a particle in plasma.
If the particle is capable of light emission-absorption, these fields become a source of spectrum alteration leading to line broadening or shift.
For example, the Holtsmark theory \cite{bib_holtsmark} predicts line broadening due to the Stark effect caused by local electric microfield.
The latter satisfies Holtsmark distribution \cite{bib_holtsmark,bib_chandra}.
Another cause of changes in spectrum can be Zeeman effect --- line splits in magnetic field.
This effect is heavily used by astronomers to measure magnetic field of stars \cite{bib_bailey,bib_mathys,bib_bailey_grunhut,bib_borra}.
But unlike electric field in Holtsmark theory, magnetic field is mostly treated as some external parameter given.

In current contribution we want to switch gears from magnetic field as an external value to the magnetic field generated by the movement of charged particles themselves.
This field may cause certain measurable effects.
For example, \cite{bib_romanovsky,bib_romanovsky_2} provide expression for probability distribution of local magnetic field and discuss its impact on Zeeman effect.
But mentioned works are based on Baranger-Moser-Iglesias theory and at some point assume particle's velocity to have normal distribution.
In current contribution we will provide alternative derivation that is independent on velocities distribution function by using Markov's method.
Presented approach is in close relation to the Chandrasekhar's derivation of the Holtsmark distribution for gravitational forces \cite{bib_chandra}.
Moreover, current approach can be utilized to estimate probability distribution of force, acting on charged particle due to other charges present in plasma.

\section{Distribution of magnetic field}
\label{sec_magn_field}

In this section we aim to calculate the probability distribution of magnetic field at the origin in an environment containing many discrete charges $\pm q$.
To begin with, let's write magnetic field due to $i$-th charge
$$
   \vec{B}_i 
 = \frac{q_i}{c r_i^3} \left[ \vec{u}_i \times \vec{r_i}\, \right],
$$
where $\vec{r}_i$, $\vec{u}_i$, $q_i$ are position, velocity, and charge respectively ($q_i = \pm q$).
The total magnetic field at the origin can be written as
$$
    \vec{B} = \sum\limits_{i = 1}^{N} \vec{B}_i,
$$
that can be treated as random walk (each $\vec{B}_i$ is a step).
Here $N$ designates the total number of particles.

Now we apply Markov's method that relates probability density function $W$ of $\vec{B}$ at the origin to its characteristic function $A_N$
\begin{subequations}
\label{eq_distr_general}
\begin{equation}
    W_{N}\left(\vec{B}\right)
 =  \frac{1}{8 \pi^3}
    \iiint\limits_{-\infty}^{+\infty}
        \exp\left(\mathbf{i} \vec{\rho} \cdot \vec{B}\right) 
        A_N\left(\vec{\rho}\,\right)
    d \vec{\rho},
\end{equation}
\begin{equation}
\label{eq_an_general}
    A_N\left(\vec{\rho}\,\right)
 =  \prod\limits_{i = 1}^{N} \iiint\limits_{\left|\vec{u}_i\right| = 0}^{c} \iiint\limits_{\left|\vec{r}_i\right| = 0}^{R}
        \exp\left(\mathbf{i} \vec{\rho} \cdot \vec{B}_i\right)
        \tau_i\left(\vec{r}_i; \vec{u}_i\right)
    d\vec{r}_i d\vec{u}_i,
\end{equation}
\end{subequations}
where $\tau_i$ is the probability distribution function for $i$-th charge.
As integration limits we chose $R$ --- radius of sphere containing $N$ charged particles and $c$ --- speed of light (because no particle can exceed it).

Now we suppose that particles are uniformly randomly distributed, i.e. the only fluctuations occurring are those compatible with a constant average density
$$
    \tau_i\left(\vec{r}_i; \vec{u}_i\right)
 =  \frac{3}{4 \pi R^3}
    \tau\left(\vec{u}_i\right)
$$
and simplify \eqref{eq_an_general} to get the following
\begin{equation}
\label{eq_an_uniform}
    A_N\left(\vec{\rho}\right)
 =  \left[
        \frac{3}{4 \pi R^3}
        \iiint\limits_{\left|\vec{u}\right| = 0}^{c} \iiint\limits_{\left|\vec{r}\right| = 0}^{R}
            \exp\left(\mathbf{i} \vec{\rho} \cdot \vec{\mathfrak{b}}\right)
            \tau\left(\vec{u}\right)
        d\vec{r} d\vec{u}
    \right]^N \!\!\!\!\!\!,
\end{equation}
where
$$
   \vec{\mathfrak{b}}
 = \frac{q}{c r^3} \left[ \vec{u} \times \vec{r}\, \right].
$$
Here we used the fact that all charges $q_i$ are equal by magnitude and may differ in sign only.
But signs $\pm$ can be dropped due to the symmetry (action of positive charge $q$ at $\vec{r}$ is equivalent to the action of $-q$ at $-\vec{r}$) and because we suppose charges to be evenly distributed (probability of finding $+q$ at some point is equal to the probability of finding there $-q$).

Now we use the following property of probability distribution
$$
    \frac{3}{4 \pi R^3}
    \iiint\limits_{\left|\vec{u}\right| = 0}^{c} \iiint\limits_{\left|\vec{r}\right| = 0}^{R}
        \tau\left(\vec{u}\right)
    d\vec{r} d\vec{u}
 =  1.
$$
Equation \eqref{eq_an_uniform} can be written as
$$
    A_N\left(\vec{\rho}\right)
 =  \left[
        1
      - \frac{3}{4 \pi R^3}
        \iiint\limits_{\left|\vec{u}\right| = 0}^{c} \iiint\limits_{\left|\vec{r}\right| = 0}^{R}
            \left[
                1
              - \exp\left(\mathbf{i} \vec{\rho} \cdot \vec{\mathfrak{b}}\right)
            \right]
            \tau\left(\vec{u}\right)
        d\vec{r} d\vec{u}
    \right]^N \!\!\!\!\!.
$$
We use the fact that $\lim_{\left|\vec{r}\right| \to \infty} \vec{\mathfrak{b}} = 0$ to conclude the inner integral converges for big $R$ and change its upper bound to infinity, but fix the particles density $n = 3 N/(4 \pi R^3)$ as $R \to \infty$ and $N \to \infty$.
Then we substitute $N = 4 n \pi R^3 / 3$ and use the definition of $e$ constant to get the final form of \eqref{eq_distr_general}
\begin{subequations}
\label{eq_general_distr}
\begin{equation}
\label{eq_W_general}
    W\left(\vec{B}\right) 
 =  \frac{1}{8 \pi^3}
    \iiint\limits_{-\infty}^{+\infty}
        \exp\left(\mathbf{i} \vec{\rho} \cdot \vec{B}\right) 
        A\left(\vec{\rho}\,\right)
    d \vec{\rho},
\end{equation}
\begin{equation}
\label{eq_A_gen}
    A\left(\vec{\rho}\,\right)
 =  \exp
    \left[
      - n
        \iiint\limits_{\left|\vec{u}\right| = 0}^{c}
            \iiint\limits_{\left|\vec{r}\right| = 0}^{\infty}
                \left[
                    1
                  - \exp\left(\mathbf{i} \vec{\rho} \cdot \vec{\mathfrak{b}}\right)
                \right]
            d\vec{r}\,
            \tau\left(\vec{u}\right)
        d\vec{u}
    \right],
\end{equation}
\begin{equation}
\label{eq_phi}
   \vec{\mathfrak{b}}
 = \frac{q}{c r^3} \left[ \vec{u} \times \vec{r}\, \right].
\end{equation}
\end{subequations}

In \eqref{eq_A_gen} integration over $\vec{r}$ can be performed.
For this we expand $\vec{\rho} \cdot \vec{\mathfrak{b}}$ and permute cyclically $\vec{\rho}$, $\vec{r}$, and $\vec{u}$ to get $\vec{r} \cdot \left[ \vec{\rho} \times \vec{u}\, \right]$.
Than we choose coordinate system so that $Z$ axis points along $\left[ \vec{\rho} \times \vec{u}\, \right]$ direction and get
$$
    A\left(\vec{\rho}\,\right)
 =  \exp
    \left[
      - n
        \iiint\limits_{\left|\vec{u}\right| = 0}^{c}
            \int\limits_{0}^{\infty} \int\limits_{-1}^{+1} \int\limits_{0}^{2\pi}
                \left[
                    1
                  - \exp\left(\frac{\mathbf{i} q}{c r^2} t \left| \vec{\rho} \times \vec{u}\, \right|\right)
                \right]
            r^2 d\varphi\, dt\, dr \,
            \tau\left(\vec{u}\right)
        d\vec{u}
    \right],
$$
where $t = \cos(\theta)$.
After integration we get
$$
    A\left(\vec{\rho}\,\right)
 =  \exp
    \left[
      - n \frac{8 \sqrt{2}}{15}
        \left( \frac{\pi q}{c} \right)^{3/2} \!\!\!
        \iiint\limits_{\left|\vec{u}\right| = 0}^{c}
            \left| \vec{\rho} \times \vec{u}\, \right|^{3/2}
            \tau\left(\vec{u}\right)
        d\vec{u}
    \right].
$$

We expect $\vec{u}$ to be uniformly distributed with respect to directions $\tau(\vec{u}) \rightarrow \tau(u)/(4 \pi u^2)$ and thus integration over angles can be performed.
This means, if length of $\vec{u}$ is fixed, it is equally probable to find it anywhere on the sphere of radius $u$ (has area $4 \pi u^2$).
We choose new coordinate system (this time $Z$ axis should point along $\rho$ direction), perform integration over $u_\varphi$ and rewrite the rest as product of integrals
$$
    A\left(\vec{\rho}\,\right)
 =  \exp
    \left[
      - n \frac{4 \sqrt{2}}{15}
        \left( \frac{\pi q \rho}{c} \right)^{3/2} \!\!\!
        \int\limits_{0}^{\pi}
            \sin^{5/2}(u_\theta)
        du_\theta
        \int\limits_{0}^{c}
            u^{3/2} \tau\left(u\right)
        du
    \right].
$$
Integration over $u_\theta$ can be performed in terms of Gamma-function $\Gamma$, while integration over $u$ yields mean value of $u^{3/2}$
$$
    A\left(\vec{\rho}\,\right)
 =  \exp
    \left[
      - n \frac{\pi \Gamma^2(-1/4)}{25}
        \left( \frac{q \rho}{c} \right)^{3/2}
        \left<u^{3/2}\right>
    \right].
$$
Further we designate $\alpha = \pi \Gamma^2(-1/4) / 25 \approx 3.02$.

Let's find expression for $W(\vec{B})$
$$
    W\left(\vec{B}\right) 
 =  \frac{1}{8 \pi^3}
    \iiint\limits_{-\infty}^{+\infty}
        \exp\left(
            \mathbf{i} \vec{\rho} \cdot \vec{B}
          - n \alpha \left( \frac{q \rho}{c} \right)^{3/2} \left<u^{3/2}\right>
        \right) 
    d \vec{\rho}.
$$
We choose $Z$ axis along $\vec{B}$ and perform integration over angles
$$
    W\left(\vec{B}\right) 
 =  \frac{1}{2 \pi^2}
    \int\limits_{0}^{+\infty}
        \frac{\sin(\rho B)}{B}
        \exp\left(
          - n \alpha \left( \frac{q \rho}{c} \right)^{3/2} \left<u^{3/2}\right>
        \right) 
    \rho d\rho.
$$
Now we want to derive an expression for $W\left(B\right)$\,--- probability density of the absolute value of $\vec{B}$.
In general, this is performed by integrating over all $\vec{B}$ having equal length.
But from the previous equation we see, that $W\left(\vec{B}\right)$ contains no dependence on angles and thus integration is reduced to simple multiplication by $4 \pi B^2$\,--- surface area of the sphere containing all $\vec{B}$ having equal absolute value.
As a result we get
\begin{equation}
\label{eq_B_dist}
    W(B)
 =  \frac{2}{\pi B}
    \int\limits_{0}^{+\infty}
        x \sin(x)
        \exp\left(
          - n \alpha \left( \frac{q x}{c B} \right)^{3/2} \left<u^{3/2}\right>
        \right) 
    dx,
\end{equation}
that has form similar to Holtsmark distribution \cite{bib_holtsmark}.
The only differences are: magnetic field $B$ is now the variable, not electric field $E$ and the second one --- coefficient in exponent is scaled down by factor $\left<u^{3/2}\right> / c^{3/2}$.
In contrast to \cite{bib_romanovsky,bib_romanovsky_2} expression \eqref{eq_B_dist} does not contain thermal velocity, but instead uses mean value of $u^{3/2}$ that is dependent on probability distribution function of velocity.

\section{Distribution of force}

Now let us calculate probability distribution of force acting on the charged particle $Q$ in an environment containing many other discrete charges $\pm q$.
To begin with, let's write down the force acting between particle, positioned at the origin and possessing velocity $\vec{v}$, and $i$-th charge of the plasma
\begin{equation}
\label{eq_force}
     \vec{F}_i 
 = - \frac{Qq_i}{r_i^3} \vec{r}_i
 +   \frac{Qq_i}{c^2 r_i^3} \left[ \vec{u}_i \times \left[ \vec{v} \times \vec{r_i}\, \right] \right],
\end{equation}
where $\vec{r}_i$, $\vec{u}_i$, $q_i$ are position, velocity, and charge respectively ($q_i = \pm q$).
The former summand in \eqref{eq_force} is the Coulomb force, while the latter is the Lorentz force.
The total force acting on the particle can be written as
$$
    \vec{F} = \sum\limits_{i = 1}^{N} \vec{F}_i,
$$
that can be treated as random walk (now $\vec{F}_i$ are random steps).
As previously, $N$ designates the total number of particles.

We apply Markov's method as we did for magnetic field $\vec{B}$ in section~\ref{sec_magn_field}.
Equations \eqref{eq_distr_general}-\eqref{eq_general_distr} and all appropriate considerations can be literally reproduced with single modification $\vec{\mathfrak{b}} \rightarrow \vec{\mathfrak{f}}$ that leads to
\begin{subequations}
\label{eq_general_distr_force}
\begin{equation}
\label{eq_W_general_force}
    W\left(\vec{F}\right) 
 =  \frac{1}{8 \pi^3}
    \int\limits_{-\infty}^{+\infty}
        \exp\left(\mathbf{i} \vec{\rho} \cdot \vec{F}\right) 
        A\left(\vec{\rho}\right)
    d \vec{\rho}
\end{equation}
\begin{equation}
\label{eq_A_inter}
    A\left(\vec{\rho}\,\right)
 =  \exp
    \left[
      - n
        \int\limits_{\left|\vec{u}\right| = 0}^{c}
            \int\limits_{\left|\vec{r}\right| = 0}^{\infty}
                \left[
                    1
                  - \exp\left(\mathbf{i} \vec{\rho} \cdot \vec{\mathfrak{f}}\right)
                \right]
            d\vec{r}\,
            \tau\left(\vec{u}\right)
        d\vec{u}
    \right]
\end{equation}
\begin{equation}
\label{eq_phi_force}
     \vec{\mathfrak{f}}
 = - \frac{Qq}{r^3} \vec{r}
 +   \frac{Qq}{c^2 r^3} \left[ \vec{u} \times \left[ \vec{v} \times \vec{r}\, \right] \right] 
\end{equation}
\end{subequations}


Integrations in \eqref{eq_general_distr_force} are quite complicated, but certain approximation may simplify it drastically.
First, we use \eqref{eq_phi_force} to express $\vec{r}$ in terms of $\vec{\mathfrak{f}}$ by rewriting vector product from \eqref{eq_phi_force} in terms of two scalar products, $\left(\vec{u} \cdot \vec{r}\,\right)$ and $\left(\vec{u} \cdot \vec{v}\,\right)$, multiplying it by $\vec{u}$ to establish $Qq \left(\vec{u} \cdot \vec{r}\,\right)/r^3 = -\left(\vec{\mathfrak{f}} \cdot \vec{u}\,\right)$, and substitute this back to \eqref{eq_phi_force}.
As a result, $\vec{r}/r^3$ can be easily obtained; squaring it we get $1/r^4$ and with this information conclude
$$
     \vec{r}
 = - \sqrt{Qq} \sqrt{c^2 + \left(\vec{u} \cdot \vec{v}\,\right)}
     \frac{c^2 \vec{\mathfrak{f}} + \left(\vec{u} \cdot \vec{\mathfrak{f}}\,\right) \vec{v}}
          {\left|c^2 \vec{\mathfrak{f}} + \left(\vec{u} \cdot \vec{\mathfrak{f}}\,\right) \vec{v}\right|^{3/2}}.
$$
Jacobian $\det J_{\vec{s}}\left(\vec{\mathfrak{f}}\,\right)$ is obtained by designating $\vec{s} = c^2 \vec{\mathfrak{f}} + \left(\vec{u} \cdot \vec{\mathfrak{f}}\,\right) \vec{v}$ and representing $\det J_{\vec{r}}\left(\vec{\mathfrak{f}}\,\right) = \det J_{\vec{r}}(\vec{s}\,) \cdot \det J_{\vec{s}}\left(\vec{\mathfrak{f}}\,\right)$.
Writing appropriate matrices and calculating their determinants we get $d\vec{s} = c^4 \left(c^2 + \left(\vec{u} \cdot \vec{v}\,\right)\right) d\vec{\mathfrak{f}}$ and $d\vec{r} = d\vec{s}~ (Qq)^{3/2} \left(c^2 + \left(\vec{u} \cdot \vec{v}\,\right)\right)^{3/2}/\left(2|\vec{s}\,|^{9/2}\right)$.
Collecting everything together \eqref{eq_A_inter} becomes
$$
    A\left(\vec{\rho}\,\right)
 =  \exp
    \left(\rule{0cm}{1cm}\right.\!\!
      - n\!\!\!\!
        \int\limits_{\left|\vec{\mathfrak{f}}\,\right| = 0}^{\infty}\!\!\!
                \left[
                    1
                  - e^{\mathbf{i} \left(\vec{\rho} \cdot \vec{\mathfrak{f}}\,\right)}
                \right]\!\!\!
            \int\limits_{\left|\vec{u}\right| = 0}^{c}
                \underbrace{
                    \frac{(Qq)^{3/2} \left(1 + \left(\vec{u}/c \cdot \vec{v}/c\right)\right)^{5/2}}
                         {2\left|\vec{\mathfrak{f}} + \left(\vec{u}/c \cdot \vec{\mathfrak{f}}\,\right) \vec{v}/c\right|^{9/2}}
                }_{\smash{I\left(\vec{v}/c\right)}}
            \tau\left(\vec{u}\right)
            d\vec{u}
        d\vec{\mathfrak{f}}\!\!
    \left.\rule{0cm}{1cm}\right).
$$
Order of integration was changed since $\vec{\mathfrak{f}}$ is now variable of integration and does not depend on anything --- all dependences  were taken into account by the Jacobian.

Now we make two assumptions.
First of all we consider a non-relativistic case, thus $v \ll c$ and $I(\vec{v}/c)$ can be approximated with its Tailor series
$$
         I\left(\vec{v}/c \approx \vec{0}\right)
 \approx I\left(\vec{0}\,\right) 
 +       \left.\left[\vec{v}/c \cdot \vec{\nabla}_{\vec{\xi}}\right] I\left(\vec{\xi}\,\right)\right|_{\vec{\xi} = \vec{0}}
 +       \left.\frac{1}{2} \left[\vec{v}/c \cdot \vec{\nabla}_{\vec{\xi}}\right]^2 I\left(\vec{\xi}\,\right) \right|_{\vec{\xi} = \vec{0}},
$$
where
\allowdisplaybreaks
\begin{align*}
    I\left(\vec{0}\,\right) 
 &= \frac{(Qq)^{3/2}}{2 \mathfrak{f}^{9/2}} \\
     \left. \left[\vec{v}/c \cdot \vec{\nabla}_{\vec{\xi}}\right] I\left(\vec{\xi}\,\right) \right|_{\vec{\xi} = \vec{0}}
 &=  \frac{(Qq)^{3/2}}{4 c^2 \mathfrak{f}^{13/2}}
     \left[
          5 \left(\vec{\mathfrak{f}} \cdot \vec{\mathfrak{f}}\,\right)  \left(\vec{v} \cdot \vec{u}\,\right)
        - 9 \left(\vec{v} \cdot \vec{\mathfrak{f}}\,\right) \left(\vec{\mathfrak{f}} \cdot \vec{u}\,\right)
     \right] \\
     \left. \frac{1}{2} \left[\vec{v}/c \cdot \vec{\nabla}_{\vec{\xi}}\right]^2 I\left(\vec{\xi}\,\right) \right|_{\vec{\xi} = \vec{0}}
 &=  \frac{3 (Qq)^{3/2}}{8\, c^4 \mathfrak{f}^{17/2}}
     \left(
           5  \mathfrak{f}^4 \left(\vec{u} \cdot \vec{v}\,\right)^2
         + 39 \left(\vec{u} \cdot \vec{\mathfrak{f}}\,\right)^2 \left(\vec{v} \cdot \vec{\mathfrak{f}}\,\right)^2 - \right. \\ &\left.
         - 6  \mathfrak{f}^2 \left(\vec{u} \cdot \vec{\mathfrak{f}}\,\right)^2 v^2
         - 30 \mathfrak{f}^2 \left(\vec{u} \cdot \vec{\mathfrak{f}}\,\right) \left(\vec{u} \cdot \vec{v}\,\right) \left(\vec{v} \cdot \vec{\mathfrak{f}}\,\right)
     \right)
\end{align*}

The second assumption is that we expect $\vec{u}$ to be uniformly distributed with respect to directions as we did previously.
It is taken into account by changing $\tau(\vec{u}\,) \rightarrow \tau(u)/(4 \pi u^2)$.
The latter means, if length of $\vec{u}$ is fixed it is equally probable to find it anywhere on the sphere of radius $u$ that has area $4 \pi u^2$.
Let us consider integral over velocities in $A(\vec{\rho}\,)$.
In spherical coordinate system it has the following form
$$
    \int\limits_{\left|\vec{u}\right| = 0}^{c}
        I\left(\vec{v}/c; \vec{u}/c\right) \tau\left(\vec{u}\,\right) d\vec{u}
  = \int\limits_0^c \int\limits_0^\pi \int\limits_0^{2\pi}
          I\left(\vec{v}/c; \vec{u}/c\right) 
         \frac{\tau(u)}{4\pi}
         \sin\left(u_\theta\right)
    du_\varphi du_\theta du.
$$
Here we have already taken into account that $\vec{u}$ is expected to be uniformly distributed with respect to directions.

Now we can integrate everything with respect to angles.
First term in Taylor series does not depend on $\vec{u}$ and thus integration over angles is equivalent to multiplication by $4 \pi$.
Integration of the second term is rather easy as well.
We integrate two summands separately choosing coordinate system so that $Z$ axis coincides with $\vec{v}$ and $\vec{\mathfrak{f}}$ respectively.
Obviously it leads to integration $\cos(u_\theta) \sin(u_\theta)$ over $[0;\pi]$ and the whole result is equal to zero.
The third term of Taylor series is integrated similarly, but instead of $\cos(u_\theta) \sin(u_\theta)$ we get $\cos^2(u_\theta) \sin(u_\theta)$ for the first three summands (is integrated to $2/3$ and $2 \pi$ comes from integration over $u_\varphi$) and the last summand can be integrated in Cartesian coordinates (relatively easily if symmetries are used).
As a result we get
$$
    \int\limits_{\left|\vec{u}\right| = 0}^{c}
        I\left(\vec{v}/c; \vec{u}/c\right) \tau\left(\vec{u}\,\right) d\vec{u}
\approx \frac{(Qq)^{3/2}}{2 \mathfrak{f}^{9/2}}
      + \frac{(Qq)^{3/2} \left<u^2\right>}{8\, c^4 \mathfrak{f}^{13/2}}
        \left(
              9 \left(\vec{v} \cdot \vec{\mathfrak{f}}\,\right)^2
            -   \mathfrak{f}^2 v^2
        \right)
      = g(\vec{v}\,).
$$
Mean value $\left<u^2\right>$ comes from the integration of $u^2$ with probability distribution function $\tau(u)$.
This is convenient result, since we can express this term through environment's temperature.

Now we perform integration over $\vec{\mathfrak{f}}$ in $A(\vec{\rho}\,)$.
We use symmetry ($g$ is invariant under $\vec{\mathfrak{f}} \rightarrow -\vec{\mathfrak{f}}$ substitution) and change $\exp\left( \vec{\rho} \cdot \vec{\mathfrak{f}} \right)$ to $\cos\left( \vec{\rho} \cdot \vec{\mathfrak{f}} \right)$, since the other part of complex exponent --- $i \sin\left( \vec{\rho} \cdot \vec{\mathfrak{f}} \right)$ --- is antisymmetric and vanishes as integration over the whole space is performed
\begin{equation}
\label{eq_A_with_ints_force}
    A\left(\vec{\rho}\,\right)
 =  \exp
    \left[
      - n
        \int\limits_{0}^{\pi} \int\limits_{0}^{\infty} \int\limits_{0}^{2\pi}
            \left[
                1
              - \cos \left( \vec{\rho} \cdot \vec{\mathfrak{f}} \right)
            \right]
        g(\vec{v}\,) \mathfrak{f}^2 \sin(\tau) d\omega d\mathfrak{f} d\tau
    \right].
\end{equation}

Integration in \eqref{eq_A_with_ints_force} is a bit tedious.
We describe a general pathway and some intermediate results without too much details.
First of all, we choose convenient coordinate system --- $Z$ axis should be directed along $\vec{\rho}$ while $Y$ axis should be perpendicular to $\vec{v}$, thus $v_y = 0$.
In this system $\vec{\mathfrak{f}}$ has coordinates $\vec{\mathfrak{f}} = \left\{\mathfrak{f} \sin(\tau) \cos(\omega); \mathfrak{f} \sin(\tau) \sin(\omega);\mathfrak{f} \cos(\tau) \right\}$.
Since $\tau$ is angle between $\vec{\mathfrak{f}}$ and $\vec{\rho}$ we can deduce $\left( \vec{\rho} \cdot \vec{\mathfrak{f}} \right) = \rho \mathfrak{f} \cos(\tau)$.
Coordinates for vector $\vec{v}$ are in this system $\vec{v} = \left\{v \sin(\nu); 0; v \cos(\nu) \right\}$.
We notice that $\left(\vec{v} \cdot \vec{\mathfrak{f}}\,\right)^2$ is the only part of integrand dependent on $\omega$ --- all the rest parts of integral are just multiplied by $2 \pi$ during integration over $\omega$.
But in a chosen coordinate system $\left(\vec{v} \cdot \vec{\mathfrak{f}}\,\right)^2$ is easily integrated as well $\int_0^{2 \pi} (\vec{v} \cdot \vec{\mathfrak{f}}\,)^2 d\omega = \pi v^2 \mathfrak{f}^2 \left( 1 - \cos^2(\tau) - \cos^2(\nu) + 3 \cos^2(\tau) \cos^2(\nu) \right)$.
Now we designate $t = \cos(\tau)$, mention that $\sin(\tau) d\tau = dt$ and $v^2 \cos^2(\nu) = \left( \vec{v} \cdot \vec{\rho} \right)^2 / \rho^2$.
This leads to
$$
\begin{aligned}
        \bar{g}(\vec{v}\,)
     &= \frac{\pi (Qq)^{3/2}}{\mathfrak{f}^{9/2}}
      + \frac{\pi (Qq)^{3/2} \left< u^2 \right>}{8\, c^4 \mathfrak{f}^{9/2} \rho^2}
        \left(
              7 v^2 \rho^2 - 9 \left( \vec{v} \cdot \vec{\rho} \right)
            + 9 t^2 \left[ 9 \left( \vec{v} \cdot \vec{\rho} \right) - v^2 \rho^2 \right]
        \right), \\
    A\left(\vec{\rho}\,\right)
&=  \exp
    \left[
      - n
        \int\limits_{0}^{\infty} \int\limits_{-1}^{1}
            \left[
                1
              - \cos \left( \rho \mathfrak{f} t \right)
            \right]
        \bar{g}(\vec{v}\,) dt\, \mathfrak{f}^2 d\mathfrak{f}
    \right].
\end{aligned}
$$
Integration over $t$ is now easily performed (doing it by parts when $t^2$ is integrated).
And as a last step everything is integrated by $\mathfrak{f}$ and we finally get
\begin{equation}
\label{eq_an_almost_final_force}
    A\left(\vec{\rho}\,\right)
 =  \exp
    \left[
      - \frac{8\sqrt{2}\, n \pi^{3/2} \rho^{3/2} (Qq)^{3/2}}{15}
        \left(
            1
          + \frac{\left<u^2\right>}{4 c^4 \rho^2}
            \left[
                v^2 \rho^2 + \left(\vec{v} \cdot \vec{\rho}\,\right)^2 
            \right]
        \right)
    \right].
\end{equation}

Minor simplification can be achieved if we suppose that $\left<u^2\right> \ll c^2$ in addition to $v^2 \ll c^2$.
This means we can approximate \eqref{eq_an_almost_final_force} with first two terms of Taylor series
\begin{equation}
\label{eq_an_finale_force}
    A\left(\vec{\rho}\,\right)
 =  e^{-a(\rho)}
    \left(
        1
      - \frac{a(\rho) \left<u^2\right>}{4 c^4 \rho^2}
        \left[
            v^2 \rho^2 + \left(\vec{v} \cdot \vec{\rho}\,\right)^2 
        \right]
    \right),
\end{equation}
where $a(\rho) = 8\sqrt{2}\, n \pi^{3/2} \rho^{3/2} (Qq)^{3/2} / 15$.
Expression \eqref{eq_an_finale_force} can be directly substituted into \eqref{eq_W_general_force}, or we may want to find its mean over all directions of $\vec{v}$ in case we are interested in microfield distribution dependent on particle's kinetic energy only (but not direction of the velocity).
In this case we get
\begin{equation}
\label{eq_w_finale_force}
    W\left(\left.F\,\right|v\right) 
 =  \frac{2}{\pi F}
    \int\limits_{0}^{+\infty}
        x \sin(x)
        e^{-a(x/F)}
            \left(
                1
              - \frac{a(x/F) \left<u^2\right> v^2}{3 c^4}
            \right)
    dx.
\end{equation}
It is worth noting that we moved from vector-valued force $\vec{F}$ to its absolute value $F$ by taking the mean value $W(F) = 4 \pi F^2 W\left(\vec{F}\right)$.

Equation \eqref{eq_w_finale_force} coincides with Holtsmark distribution \cite{bib_holtsmark} if $\langle u^2 \rangle = 0$.
The latter means Holtsmark distribution is special case of \eqref{eq_w_finale_force} when temperature is zero ($m \langle u^2 \rangle \sim kT$).

\section{Results}

We have obtained stationary probability distribution functions for absolute value of magnetic field $B$ \eqref{eq_B_dist} and force $F$ \eqref{eq_w_finale_force} acting on charged particle that moves with velocity $v$ in plasma environment.
Let us briefly analyze these findings.

\begin{figure}[ht]
\includegraphics[width=\textwidth]{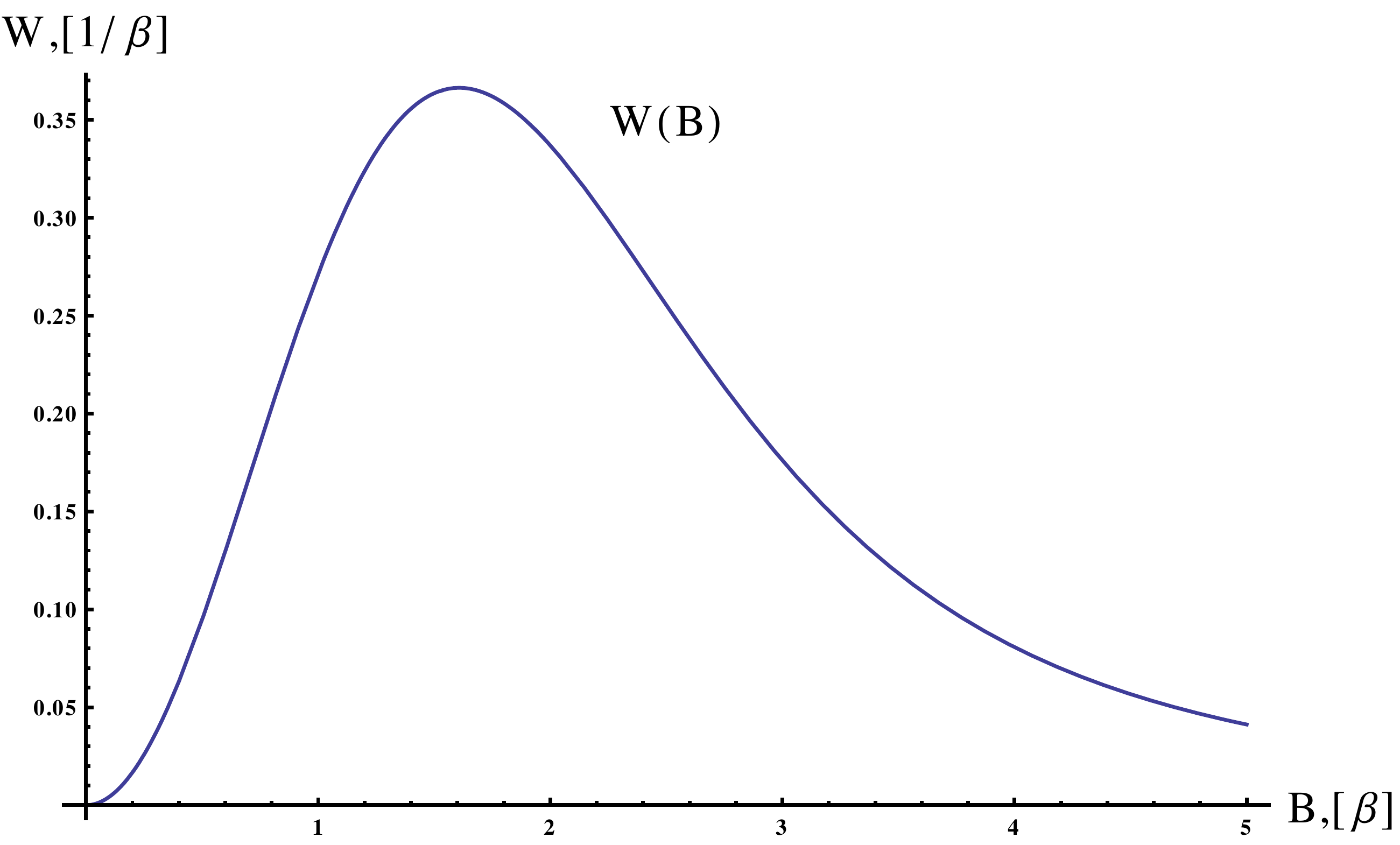} 
\caption{Probability distribution function $W(B)$ in terms of characteristic magnetic field $\beta$.}
\label{fig_w_plot}
\end{figure}

First of all, it is convenient to rewrite \eqref{eq_B_dist} in a bit different form.
We introduce variable
\begin{equation}
\label{eq_beta}
    \beta = n^{2/3} \alpha^{2/3} q \frac{\langle u^{3/2} \rangle^{2/3}}{c},
\end{equation}
that has dimension of $[B]$ and thus is characteristic magnetic field of the system.
Now we can measure magnetic field $B$ and the probability distribution $W(B)$ in terms of characteristic field $\beta$, e.g. $B = \beta b$ and $W(B) = w(b) / \beta$, that leads to
\begin{equation}
\label{eq_holtsmark}
    w(b)
 =  \frac{2 b}{\pi}
    \int\limits_{0}^{+\infty}
        x \sin(b x)
        \exp\left(-x^{3/2}\right) 
    dx.
\end{equation}
The plot of \eqref{eq_holtsmark} looks as in figure~\ref{fig_w_plot}.
This is Holtsmark distribution; same distribution have gravitational field of randomly arranged masses \cite{bib_chandra} and electrostatic field of randomly placed charges (as in plasma) \cite{bib_holtsmark}.
Thus, in some sense, we could call this result expectable.

One should notice, equation \eqref{eq_beta} contains $\langle u^{3/2} \rangle$ without specifying how this number should be obtained.
The reason is, we did no assumptions on velocity distribution function up to this point.
This is in contrast to theories based on, for example, Baranger-Moser-Iglesias theory \cite{bib_romanovsky,bib_romanovsky_2} that assume particle's velocity to have normal distribution.
Thus, obtained equations should be applicable to broader spectrum of models.

For practical use we may estimate few valuable parameters by means of numeric calculations
\begin{center}
\begin{tabular}{cccc}
Most probable $B$     & Mean, $\langle B \rangle$   & $\sqrt{\langle B^2 \rangle}$    & $\sqrt{\langle B^2 \rangle - \langle B \rangle^2}$ \\
\cmidrule(lr){1-1}      \cmidrule(lr){2-2}            \cmidrule(lr){3-3}                \cmidrule(lr){4-4}
$\approx 1.6\, \beta$ & $\approx 3.2 \, \beta$      & $\approx 6.9 \, \beta$          & $\approx 6.1 \, \beta$                               \\
\end{tabular}
\end{center}
We see that $B$'s order of magnitude is approximately equal to the characteristic magnetic field $\beta$ \eqref{eq_beta}.
As an immediate consequence, we roughly estimate Zeeman effect for hydrogen molecule
\begin{equation}
\label{eq_zeeman}
            \Delta E 
    \sim    \frac{2 \hbar e}{m_e c} \langle B \rangle
    \approx 17.5 \times 10^{-9} \frac{\text{eV}}{\text{Gauss}} ~ \beta \text{[Gauss]}.
\end{equation}
We expect this effect to be measurable in some cases.

Now let us switch gears to the force acting on charge.
We have already obtained probability distribution function \eqref{eq_w_finale_force}, but it still can be refined if we measure everything in terms of some characteristic force $\phi$ and constant $\omega$ comprising velocities
\begin{equation}
\label{eq_char_force}
    \phi = \sqrt[3]{\frac{128}{225}}\, \pi^2 n^{2/3} Qq;\quad \omega = \frac{\left<u^2\right> v^2}{3 c^4}.
\end{equation}
We may suppose $W(F|v) = w(f|\omega) / \phi$ and $F = \phi f$ thus getting
\begin{equation}
\label{eq_dist_force}
    w(f|\omega)
 =  \frac{2 f}{\pi}
    \int\limits_{0}^{+\infty}
        x \sin(f x)
        \exp\left(-x^{3/2}\right)
        (1 - x \omega)
    dx.
\end{equation}

We see that \eqref{eq_dist_force} is very similar to the \eqref{eq_holtsmark} (Holtsmark distribution).
The only difference seen is correction term $x \omega$.
From definition of $\omega$ \eqref{eq_char_force} we know that $0 < \omega < 1/3$, but due to assumptions that velocities in the system are not relativistic, we can assume $\omega \ll 1/3$.
In case of $\omega = 0$ (e.g. zero temperature) we get exactly the Holtsmark distribution.

Let us provide some estimations as we did for magnetic field $B$.
\begin{center}
\begin{tabular}{cccc}
Most probable $F$                  & Mean, $\langle F \rangle$    & $\sqrt{\langle F^2 \rangle}$    & $\sqrt{\langle F^2 \rangle - \langle F \rangle^2}$ \\
\cmidrule(lr){1-1}                   \cmidrule(lr){2-2}            \cmidrule(lr){3-3}                \cmidrule(lr){4-4}
$\approx (1.6 + 0.6\, \omega)\phi$ & $\approx (3.2 + 5.1\, \omega)\phi$ & $\approx (6.9 + 9.0\, \omega)\phi$ & $\approx (6.1 + 7.0\, \omega)\phi$                               \\
\end{tabular}
\end{center}
All results in this table were obtained as approximations for small $\omega$, expanding expressions into Tailor series where needed.
From the table we see that $F$'s order of magnitude is equal to characteristic force $\phi$.
In contrast to similar result for $B$, proportionality coefficient is not strictly constant, but contains small correction proportional to $\omega$.
We suppose that studies of Brownian particles in dusty plasma may provide some experimental verification of provided results.

\section*{Conclusion}

In conclusion, we have obtained stationary probability distribution $W(B)$ \eqref{eq_B_dist} of magnetic field $B$ that is generated by movement of charges in plasma and stationary probability distribution of force $W\left(\left.F\,\right|v\right)$ \eqref{eq_w_finale_force} acting on charged particle that moves with velocity $v$ in plasma environment (electrostatic and magnetic interactions are taken into account).
Introducing characteristic magnetic field $\beta$ \eqref{eq_beta} and characteristic force $\phi$ \eqref{eq_char_force} we obtained two dimensionless distributions -- $w(b)$ \eqref{eq_holtsmark} and $w(f|\omega)$ \eqref{eq_dist_force}.
The former \eqref{eq_holtsmark} perfectly coincides with Holtsmark distribution and looks similar to the one obtained in \cite{bib_romanovsky,bib_romanovsky_2}.
But in contrast to these works we neither assumed any particular distribution of the velocities nor used Baranger-Moser-Iglesias theory and thus \eqref{eq_beta} contains mean value $\langle u^{3/2} \rangle$ that is dependent on unspecified velocities distribution making it compatible with wider spectrum of models.
This result was in some sense expected since gravitational field of randomly arranged masses \cite{bib_chandra} and electrostatic field of randomly placed charges (as in plasma) \cite{bib_holtsmark} manifest same distribution.
We suppose, obtained result can be verified experimentally, e.g. by means of Zeeman effect we roughly estimated \eqref{eq_zeeman}.

Distribution \eqref{eq_dist_force} is a modification of Holtsmark distribution -- it contains purely Holtsmark part that corresponds to electrostatic interaction and correction proportional to $\omega$ \eqref{eq_char_force} that corresponds to magnetic interaction.
At zero temperature, i.e. $\langle u^2 \rangle = 0 \rightarrow \omega = 0$, obtained distribution perfectly coincides with Holtsmark distribution.
One should notice, deriving \eqref{eq_w_finale_force} we used the fact that velocities $\vec{u}_i$ of moving charges are uniformly distributed with respect to directions, thus velocity $v$ should be measured with respect to a frame of reference possessing this property.
We suppose, obtained results may be valuable to studies of Brownian motion in plasma environment.

\vspace{1cm}\noindent
\textbf{Acknowledgement:} funding for this research was provided by the State Fund for Fundamental Research of Ukraine (Project №~F~76/84).



\end{document}